# Temperature Studies for ATLAS MDT BOS Chambers

A. Engl, R. Hertenberger, O. Biebel, R. Mameghani, D. Merkl, F. Rauscher, D. Schaile and R. Ströhmer

*Abstract*—Data sets with high statistics taken at the cosmic ray facility, equipped with 3 ATLAS BOS MDT chambers, in Garching (Munich) have been used to study temperature and pressure effects on gas gain and drifttime. The deformation of a thermally expanded chamber was reconstructed using the internal RasNik alignment monitoring system and the tracks from cosmic data. For these studies a heating system was designed to increase the temperature of the middle chamber by up to 20 Kelvins over room temperature. For comparison the temperature effects on gas properties have been simulated with Garfield. The maximum drifttime decreased under temperature raise by $-2.21 \pm 0.08\ \frac{ns}{K}$, in agreement with the results of pressure variations and the Garfield simulation. The increased temperatures led to a linear increase of the gas gain of about $2.1\%\ \frac{1}{K}$.

The chamber deformation has been analyzed with the help of reconstructed tracks. By the comparison of the tracks through the reference chambers with these through the test chamber the thermal expansion has been reconstructed and the result shows agreement with the theoretical expansion coefficient. As the wires are fixed at the end of the chamber, the wire position calculation can not provide a conclusion for the chamber middle. The complete deformation has been identified with the analysis of the monitoring system RasNik, whose measured values have shown a homogeneous expansion of the whole chamber, overlayed by a shift and a rotation of the chamber middle with respect to the outer part of the chamber. The established results of both methods are in agreement. We present as well a model for the position-drifttime correction as function of temperature.

## I. Introduction

GRADIENTS of temperature in the ATLAS detector influence the resolution of single Monitored Drift Tube (MDT) chambers[1], [2]. Due to heat from the inner part of the ATLAS detector and from the electronics of the muon spectrometer, temperature differences of about 3 to 5 Kelvins are expected along a single Barrel Outer Small (BOS) chamber.

To study the effect of temperature on driftgas properties and possible deformations of the chamber itself and to provide a correction of the space drift time relation, a thermally isolated BOS MDT chamber was heated up with warm air.

## II. Setup

### A. MDT Chamber

An ATLAS MDT chamber (see Fig. 1) consists of 2 multilayers with 3 or 4 layers of 30 mm drift tubes each[3]. The tubes are filled with an Argon and $CO_2$ gas mixture of the ratio 93:7 at an absolute pressure of 3 bar. The voltage at the gold plated W-Re anode wire (diameter 50 $\mu m$) is 3080 V. These conditions correspond to a maximum electron drift time of 700 ns and a gas gain of $2 \cdot 10^4$. The tubes are glued onto a support structure (2 long bars, 3 crossplates). On the HV side high voltage and gas services are provided to the chamber, whereas on the other side the signals are read out (RO).

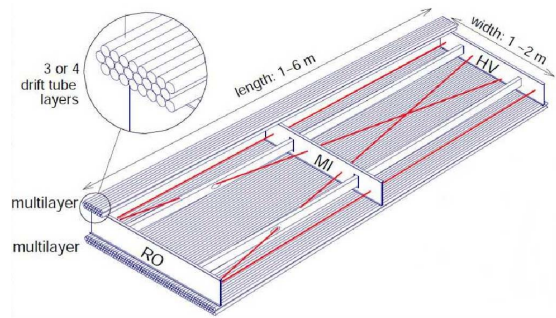

Fig. 1. Technical scheme of an ATLAS MDT chamber. Each red line corresponds to a light beam of the RasNik system.

### B. Cosmic Ray Facility

The measurements were performed at the cosmic ray facility in Garching (Fig. 2). A set of three MDT chambers, 2 reference chambers and in between the test chamber, is enclosed by layers of scintillation counters on the top and bottom to trigger for cosmic muons by a coincidence of the 2 signals. The trigger rate was about 100 Hz. An 40 cm iron absorber allowed us to select muons with an energy higher than 600 MeV to reduce multiple scattering. The

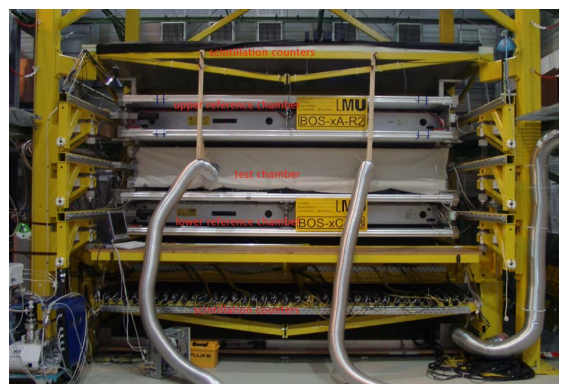

Fig. 2. Cosmic ray facility in Garching (Munich).

Ludwig-Maximilians University Munich, LS Schaile, Am Coulombwall 1, 85748 Garching; contact: Albert.Engl@physik.uni-muenchen.de

MDT chambers were operated under the above described ATLAS conditions and the ambient temperature was $20\,°C$. Using an electronically controlled outlet valve, stable pressure conditions were guaranteed. At increased chamber temperature and stabilized pressure, therefore the gas density decreases.

*C. Heating Setup*

To achieve an almost homogeneous temperature increase of the test chamber, the heating system shown in (Fig. 3) was developed. The warm air of two fan heaters (P = 2 kW) is brought into the chambers space via flexible and isolated tubes. A microperforated tube, lying on the top of the lower multilayer inside the chamber, guaranteed that the warm air spreaded homogeneously over the whole chamber. Thermal isolation of the test chamber provided by polystyrol panels and cotton rags guaranteed that the reference chambers remained thermally uninfluenced during the measurement periods. 18 temperature sensors, integrated in a BOS chamber[3], were used to observe the temperature during the studies. It was possible to heat the test chamber up to $40\,°C$, $20K$ above room temperature, with a stability of $\pm1.5K$ during a measurement period of 5 days.

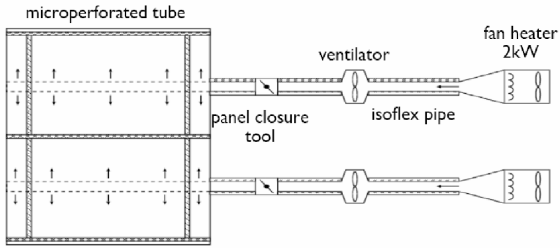

Fig. 3. Scheme of heating setup.

*D. The Optical Monitoring System*

The RasNik[4] system was developed by NIKHEF to monitor changes in the chamber geometry (MDT = Monitored Drift Tube). An optical mask on the HV side of the BOS chamber is illuminated by infrared LEDs. Its image is projected by lenses positioned in the middle of the chamber (MI) onto a CCD sensor at the readout side (RO) (see Fig. 4). Deviations of about 2 $\mu m$ can be reliably measured by this system. Each chamber is equipped with 4 RasNik devices, shown in Fig. 5. This constellation provides the possibility to distinguish between expansions of RO-, HV- and MI-crossplate, providing together with two longbars the skeletal structure of a chamber. The shift of a single crossplate can also be observed. Typical signatures of changes in geometry are listed in Tab. 1.

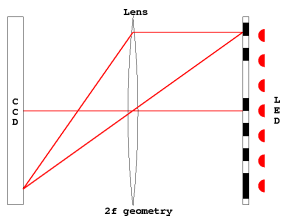

Fig. 4. The parts of the RasNik system.

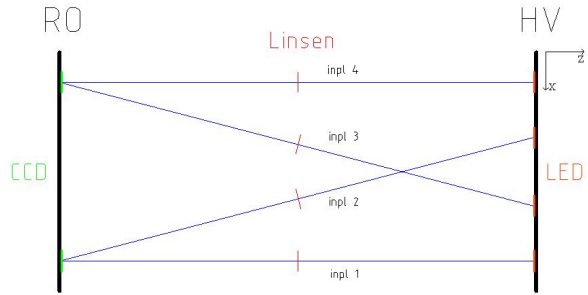

Fig. 5. Geometry of the 4 RasNik devices, named inpl 1 to inpl 4, on a chamber.

| | |
|---|---|
| $X_{1,2,3,4} = \|X\|$ $sgn(X_{1,2,3,4}) = \pm 1$ | shift of a crossplate |
| $X_{1,2,3,4} = \|X\|$ $sgn(X_{1,2}) = +1$ $sgn(X_{3,4}) = -1$ | expansion of RO-crossplate |
| $X_{1,4} = \|X\|; X_{2,3} = \frac{1}{3}\|X\|$ $sgn(X_{1,2}) = +1$ $sgn(X_{3,4}) = -1$ | expansion of MI-crossplate |
| $X_{1,4} = \|X\|; X_{2,3} = \frac{1}{3}\|X\|$ $sgn(X_{1,3}) = +1$ $sgn(X_{2,4}) = -1$ | expansion of HV-crossplate |
| $X_{1,4} = 0; X_2 = X;$ $X_3 = -X$ | trapezoidal expansion |

Tab. 1. The signatures of the reconstructable changes in geometry of the chamber.

III. DEFORMATION OF THE CHAMBER

*A. Analysis of the RasNik System*

Fig. 6 shows that all measured RasNik deviations in X direction show a linear behaviour with increasing temperature, so the shape of the thermally caused deformation stays constant in this temperature range, but the amplitude of the deformation increases. Considering the X values, one can see that all deviations are positive but have different amplitudes. These values indicate a shift of the MI-crossplate (see Tab. 1) superimposed by different thermal expansion of the three crossplates. The measured deviations in Y direction (see Fig 6) show linear increase of the amplitudes at each measured temperature and therefore correspond to a rotation of the MI-crossplate with respect to RO and HV sides.
The average expansion agrees with the thermal expansion coefficient of aluminium of $24 \cdot 10^{-6} \frac{1}{K}$.

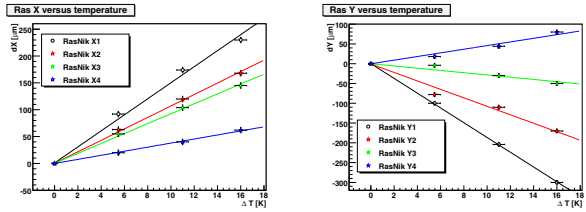

Fig. 6. Deviations in X direction (left) and Y direction (right) of the 4 RasNik systems in the chamber.

## B. Analysis Using Cosmic Muon Tracks

Another way to determine the expansion of RO- and HV-crossplate was the measurement of the wire positions using cosmic muons[5]. The tracks of the reference chambers are compared to those of the test chamber. The transformations[5] needed to obtain aligned track segments yield an average expansion which is in agreement with the RasNiks result. Fig. 7 shows the result of the wire position determination after the correction of the average expansion. One can see that the crosses (red) differ from the nominal position (black circles) and show decreasing behaviour. This implies that a thermal expansion with different amplitudes for each crossplate took place. This analysis confirmed the results of the RasNik analysis.
The observed deformation corresponds to a banana-like bended chamber (Fig. 8), where the middle of the chamber is shifted and rotated.

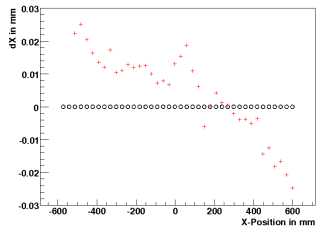

Fig. 7. Nominal (black circles) and measured (red crosses) wire positions after heating by $\Delta T = 12\ K$.

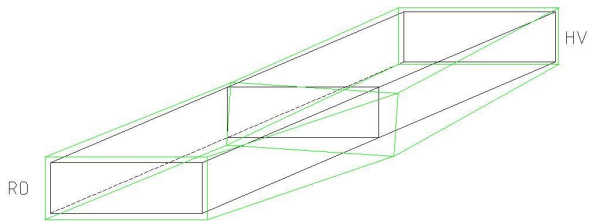

Fig. 8. Scheme of the chamber geometry before (black) and after heating (green) by $\Delta T = 12\ K$.

## IV. GAS PROPERTIES AT DIFFERENT TEMPERATURES

### A. Maximum Drift Time

The maximum drift time $t_{max}$ is defined as the difference between leading and trailing edge of the drift time spectrum (Fig. 9). To obtain $T_0$ and $T_{max}$, 2 Fermi type functions are fitted at the beginning and the end of the drift time spectrum. Fig. 7 shows that $t_{max}$ changes linearly with the temperature by $-2.21 \pm 0.08\ \frac{ns}{K}$. It changes similarly if pressure is reduced to get the same gas density as for a temperature change. Both variations show the same influence on the drift time and are caused predominantly by the density fluctuation. Garfield[6] simulations (not shown in Fig 10) agree with this result.

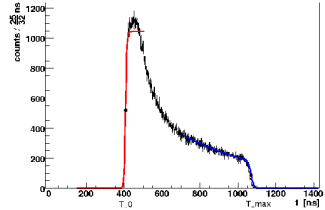

Fig. 9. Drift time spectrum including fit functions.

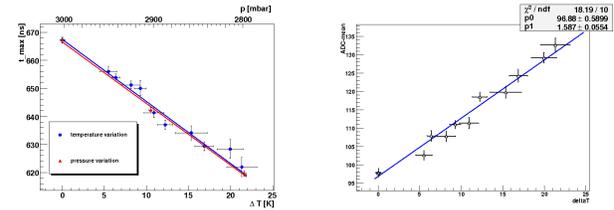

Fig. 10. Change of max. drift time with temperature and pressure normalised by constant density (left) and change of ADC mean with temperature (right).

### B. Gas Gain

To monitor the change in the gas gain, we used the mean value of the Analog to Digital Converter (ADC) signal which is proportional to the gas gain. The gas gain showed an increase of 2.1 % per Kelvin with temperature increase (see Fig. 10).

## V. TEMPERATURE PARAMETRIZATION OF THE rt-RELATION

To avoid systematic errors on the position measurement of muons, the position-drift time relation (r-t relation) must be well determined. Taking 20 sampling points, each at constant drift time, and interpolating between the points leads to a precise r-t relation. With increasing temperature the deviation of every sampling point $dr$ shows linear behaviour (Fig. 11). The gained slope time relation can be described with a correction function $C(t)$ and so the change of the drift radius $dr$ with the temperature is given by $C(t)$ multiplied with the temperature change.

$$dr(T,t) = C(t) \cdot \Delta T \quad (1)$$

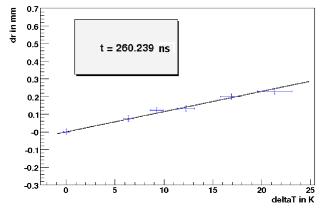

Fig. 11. Change of drift radius $dr$ with the temperature at the sampling point t = 260 ns.

As Fig. 12 shows, the calculated change of the drift radius agrees with the measurement.

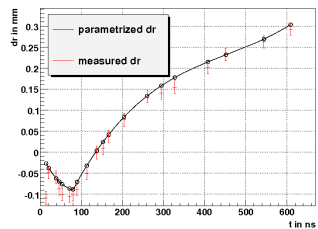

Fig. 12. Comparison of measured $dr$ (red) and parametrized $dr$ (black) at $\Delta T = 12\ K$.